\documentclass[prl,twocolumn,showpacs,amsmath,amssymb,superscriptaddress]{revtex4}

\usepackage{graphicx}
\usepackage{dcolumn}
\usepackage{bm}

\usepackage[latin1]{inputenc}

\newcommand{\bk}{\mathbf{k}}

\renewcommand\Re{\operatorname{Re}}
\renewcommand\Im{\operatorname{Im}}

\begin{document}

\title{Evidence of transient interactions between $\pi\rightarrow\pi^*$ optical excitations and image potential states in graphite} 

\author{M. Montagnese}
\author{S. Pagliara}
\affiliation{Dipartimento di Matematica e Fisica, Università Cattolica del Sacro Cuore I-25121 Brescia, Italy}
\author{S. Dal Conte}
\affiliation{Dipartimento di Fisica "A. Volta", Università degli Studi di Pavia I-27100 Pavia, Italy}
\author{G. Galimberti}
\author{G. Ferrini}
\affiliation{Dipartimento di Matematica e Fisica, Università Cattolica del Sacro Cuore I-25121 Brescia, Italy}
\author{F. Parmigiani}
\email{fulvio.parmigiani@elettra.trieste.it}
\affiliation{Dipartimento di Fisica, Università degli Studi di Trieste I-34127 Trieste, Italy \\and Sincrotrone Trieste S.C.p.A. I-34012 Basovizza, Italy}

\date{\today}

\begin{abstract}
Here we report the experimental evidence of the interactions between the excitations of the $\pi\rightarrow\pi^*$ optical transition and the image potential states (IPS) of highly oriented pyrolitic graphite (HOPG). By using non-linear angle resolved photoelectron spectroscopy (NL-ARPES) we show that the IPS photoemission intensity, the effective mass, and the linewidth exhibit a strong variation when the photon energy is tuned across the $\pi\rightarrow\pi^*$ saddle points in the 3.1 - 4.5 eV photon energy range. A model based on the self-energy formalism is proposed to correlate the effective mass and the linewidth variations to transient many body effects, when a high carriers density (in the $10^{20}$ cm$^{-3}$ range) is created by the absorption of a coherent light pulse. This finding brings a clear evidence of a high IPS-bulk coupling in graphite and opens the way for exploiting the IPS as a sensitive, nonperturbing probe for the many-body dynamics in materials.
\end{abstract}

\pacs{81.05.uf - 73.20.-r - 79.60.-i - 71.20.Gj}

\maketitle

The concept of quasiparticle in solids has been very useful for describing many unconventional physical properties, such as the deviation from the linear response in many body systems, by taking into account the residual interactions between the quasiparticles, i.e. the part of the Coulomb forces not accounted for in the formation of quasiparticle \cite{ChemlaNature2001}. 

For example, in these last decades it has been demonstrated that the optical properties of semiconductors near the electronic bandgap are dominated by excitons and their residual interactions, producing many-body effects, such as band-gap renormalization (BGR) \cite{BerggrenPRB1981}, and non-linear optical mechanisms. In particular, a deep insight has been gained on understanding the BGR in semiconductors by means of time-resolved non-linear optical spectroscopy in the fs time-domain \cite{ChemlaSS1999}. For these reasons and also because their linear properties are well understood and excitons can be easily created and controlled through band-gap optical transitions, semiconductors have been considered ideal materials for such studies. Instead, in semi-metallic and gap-less systems the response to photoexcitation processes is still unclear, and the studies of the dynamics and interactions of quasiparticles have been only marginal.

In a previous work \cite{PagliaraSS2008}, we have shown that for nonlinear photoemission processes in metals, the IPS effective mass ($m^{*}$) can be influenced by the hot electron (hot-e$^{-}$) gas generated during the excitation process, but the interplay between bulk many body excitations and surface states remains elusive. Nonetheless, understanding these processes is of paramount importance for condensed matter physics. 

In this framework, graphite is an intriguing and interesting system, since the excitations in this material could have a strong coupling with the IPS, originating from the hybridization between the IPS and the interlayer band \cite{PosternakPRL1983}, along with the anisotropy of the charge transport in the direction normal to the graphene planes. In addition, the presence of a van Hove singularity (vHs) at $\hbar\omega=4.4$ eV in the JDOS, corresponding to the $\pi \rightarrow \pi^*$ transition between the bands saddle points, located at the M point of the Brillouin Zone (BZ) \cite{BassaniNC1966}, allows the photoinjection of high carrier densities in the outermost graphene layers.  

Here we report on angle-resolved non-linear photoemission spectroscopy (NL-ARPES) studies, using ultrashort coherent photon pulses, on HOPG.
By tuning the photon energy in the 3.1-4.5 eV range we discover that the photoemission (PE) intensity of the n=1 IPS, populated by the hot-e$^-$ scattering process accompanying the $\pi \rightarrow \pi^*$ femtosecond excitations, has a maximum near $\hbar\omega=4.0$ eV. This energy results $\approx0.4$ eV lower than the maximum of the imaginary part of the dielectric function, $\epsilon_2(\omega)$, where the highest photoelectron yield is expected for indirectly populated IPS. 

Stimulated by this result and its possible link to transient BGR effects, we measured also the IPS effective mass ($m^*$) and linewidth ($\gamma_\mathrm{tot}$), finding that both quantities reach a maximum between $\hbar\omega=4.0$ eV and $\hbar\omega=4.1$ eV. On the basis of the measured $m^*$ and $\gamma_\mathrm{tot}$ versus the photon energy we craft a semi-quantitative self-energy model, allowing to obtain an estimation of the renormalization energy of $\approx-0.4$ eV. This value is fully consistent with a BGR mechanism that, by downward shifting the vHs energy, sets the IPS PE maximum in the $\hbar\omega\approx4.0$ eV range, in agreement with the experimental observations. 

\begin{figure}[htb]
\begin{center}
\includegraphics[width=0.95\linewidth, bb=67 262 503 664]{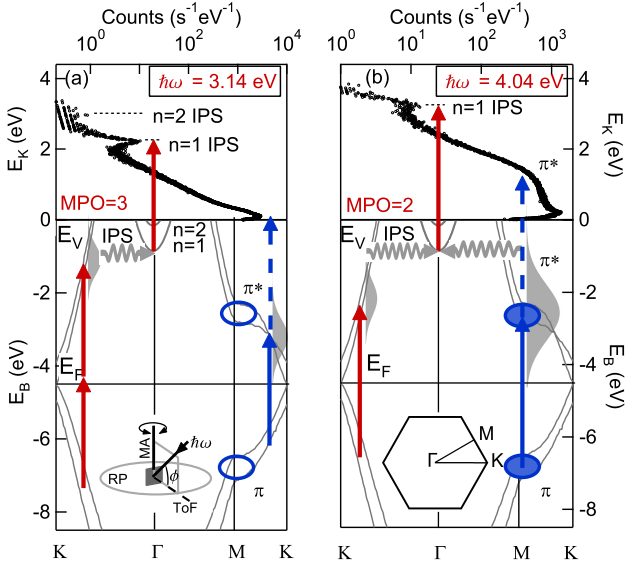}
\caption{(a) IPS population-PE process (red) and $\pi$ band transition (blue) at $\hbar\omega$=3.14 eV (far from the vHs resonance, indicated by the empty blue ellipses). Solid (dashed) lines represents direct (indirect) photoinduced transitions. Gray wavy lines represent momentum-conserving scattering. Gray-filled, bell-like curves represent hot-e$^-$ distributions; a sketch of the symmetric experimental geometry is shown. The incident pulse is steered onto the sample at an azimuthal angle $\phi=45$° with respect to the rotation plane (RP) and in-plane with the ToF and manipulator (MA) axis. (b) The same as for (a) at $\hbar\omega$=4.04 eV (near the vHs resonance, full blue ellipses). The contribution to the IPS population from hot-e$^-$ scattering from the M point is shown; the $k_z=0$ BZ is shown in the lower part of the figure; ($\bk$ is the crystal momentum). Bulk graphite bandstructure is taken from Ref. \onlinecite{TatarPRB1982}; E$_F$ and E$_V$ are the Fermi and vacuum energies, respectively.}
\label{FIG1}
\end{center}
\end{figure}

The experimental setup is based on an amplified Ti:Sapphire laser system and optical parametric amplifiers, producing p-polarized, $\approx 100$ fs, near-UV pulses that are focused on the HOPG sample kept in an ultra-high vacuum chamber (pressure $<2.0\cdot10^{-10}$ mbar). Photoelectrons are detected by a custom-built time of flight (ToF) electron spectrometer \cite{PaolicelliSRL2002} with an angular acceptance of $\pm0.85$° and an overall energy resolution 35 meV at an electron kinetic energy ($E_K$) of $\approx2$ eV.  The HOPG samples are cleaved \textit{ex-situ} and annealed at 450°C until a good LEED pattern is obtained. The photoelectron spectra have been collected either at $\bk_{||}=0$ (normal emission - NE) and $\bk_{||}\neq 0$, being $\bk_{||}$ the component of the momentum parallel to the surface, by varying the angle between the sample normal and the ToF axis. In order to perform dispersion measurements with the maximal axial symmetry possible, the plane of incidence of the incoming UV pulses has been chosen to lie, at NE, on the plane defined by the manipulator axis and the ToF axis. [see lower part of Fig. \ref{FIG1}-(a)]. All measurements have been performed at room temperature.

The upper parts of Fig. \ref{FIG1} show normal emission NL-PE spectra of HOPG at $\hbar\omega=3.14$ eV (a) and $\hbar\omega=4.04$ eV (b). The peaks located at E$_K$ above $2$ eV are identified with the n=1 and n=2 IPS emission, instead the hump-shaped feature ($\pi$*), located below $2$ eV in Fig \ref{FIG1} (b), is assigned to the linear PE from the $\pi^{*}$ band, transiently populated via the $\pi\rightarrow\pi^*$ transition. These assignments are consistent with the s-polarized photon spectra (not shown here) where the IPS emissions are quenched and the $\pi^*$ intensity is significantly reduced, according to the electric dipole selection rules for IPS and bulk bands with $\pi$ symmetry \cite{hufner}. 

To properly describe these complicated non-linear processes, important insights can be achieved by measuring the PE multiphotonic order (MPO) at different photon energies, defined as the logarithmic slope of the IPS emission peak intensity versus the pump fluence. We find a MPO=3 for $\hbar\omega=3.14$ eV and MPO=2 for $\hbar\omega=4.04$ eV. These informations, along with the graphite bandstructure \cite{TatarPRB1982}, reported in the lower parts of Fig. \ref{FIG1}, shows that, at $\hbar\omega=3.14$ eV, the IPS population process is the result of a two-photon absorption mechanism, whereas at $\hbar\omega$=4.04 eV the IPS is populated by one-photon absorption process. At both photon energies the IPS is indirectly populated trough electron scattering of the hot electron (hot-e$^-$) population created in the $\pi^*$ band by the optical absorption (see Fig. \ref{FIG1}).

\begin{figure}
\begin{center}
\includegraphics[width=0.9\linewidth, bb=41 11 538 504]{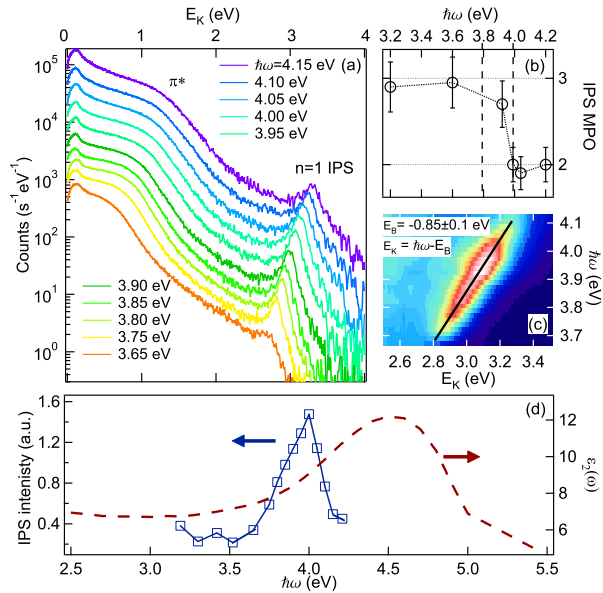}
\caption{(a) NL-PE spectra of HOPG. A selection of spectra collected with photon energies between 3.65 eV and 4.15 eV is shown here. (b) n=1 IPS MPO versus photon energy, showing a MPO=3 to MPO=2 transition in the 3.8-4.0 eV region. (c) The IPS region of the NL-PE spectra shown in (a) versus photon energy in a false-color representation. Black line is a linear fit of the IPS position versus photon energy. (d) IPS intensity (blue squares) versus photon energy. The solid line is a guide to the eye. The red dashed line represents the graphite $\epsilon_2$ taken from ref. \onlinecite{TaftPR1965}.}
\label{FIG2}
\end{center}
\end{figure}

To unambiguously clarify the IPS population mechanism several NE PE spectra, a selection of which is shown in \ref{FIG2}(a), have been measured at a constant laser fluence F = 150 $\mu$J cm$^{-2}$ by tuning the photon energy from $\hbar\omega$ = 3.2 eV to $\hbar\omega$ = 4.2 eV. 
The MPO of the IPS photoemission, versus photon energy, are shown in Fig. \ref{FIG2}(b). As it is possible to note the MPO changes from 3 to 2 in going form 3.2 eV to 4.15 eV photon energy, being the transition region $\approx300$ meV-wide and centered at $\hbar\omega=3.9$ eV. 
The linearity of the photoemission process from the IPS is shown in Fig. \ref{FIG2}(c) where the IPS emission intensity is reported in a false colors plot in the ($E_K,\hbar\omega$) plane. It is manifest that the relation between the $E_K$ values of the IPS maxima and the photon energy is linear and it is well fitted by $E_K=\hbar\omega+E_B$ (black line), being $E_B$ the IPS binding energy referred to the vacuum level, $E_V$. Noteworthy, the IPS $E_B$ value obtained from the data fit, i.e. $E_B=-0.85\pm0.1$ eV is in agreement with the only value found in literature \cite{LehmannPRB1999}. 
These findings confirm that at $\hbar\omega<3.8$ eV an indirect two-photon population process plus a linear photoemission process (2+1=3 MPO) take place, whereas at $\hbar\omega>4$ eV the IPS photoemission spectra result from an indirect one-photon population plus a linear photoemission process (1+1=2 MPO). For further details, see Ref. \onlinecite{Montagnese_unp}.

Being the population process of the IPS the result of scattering of the hot-e$^-$ gas excited into the $\pi^*$ bands, the IPS emission intensity should be proportional to the number of optically excited carriers in the $\pi^*$ band when the photon energy is tuned through the $\pi\rightarrow\pi^*$ transition, i.e. proportional to $\epsilon_2(\omega)$. From the NE PE spectra is possible to obtain the dependence of the IPS emission intensity versus the photon energy. Fig. \ref{FIG2}(d) shows that the IPS photoemission intensity depends on the photon energy, presenting a maximum at $\hbar\omega\approx4.0$ eV. However, the $\epsilon_2(\omega)$, as taken from Ref. \onlinecite{TaftPR1965}, has a maximum, resulting from the $\pi\rightarrow\pi^*$ optical transition between the two saddle points located at the M point of the Brillouin zone (BZ), i.e. at the vHs, located at $\hbar\omega=4.4$ eV. From these data it is unambiguous that the IPS maximum is located $\approx0.4$ eV below the $\epsilon_2(\omega)$ maximum. The MPO changes, along with the graphite band structure, cannot explain this behavior, and, in particular, the sudden drop in the PE yield detected at $\hbar\omega>4$ eV. Moreover, no enhancements of the linear photoemission, due to final state effects or other band structure mechanisms, are expected in the 4 eV range, from bandstructure analysis \cite{TatarPRB1982}. 

A possible explanation for this discrepancy can be gained by considering a transient shrinking of the gap between the two saddle points originating from the high density of photoinduced carriers, which at an absorbed fluence of $(1-R)F\approx 100$ $\mu$J cm$^{-2}$, are $\approx2\cdot10^{20}$ cm$^{-3}$, i.e. two orders of magnitude greater than the room temperature carrier density at the Fermi level in graphite \cite{McClurePR1956}. Eq. \ref{nom} has been used to estimate carrier densities, and a constant reflectivity $R=0.3$ for the 3.2-4.5 eV range \cite{GreenwayPR1969} has been considered.
It is worth noting that these excitations are in the near-UV region and across the non-dispersive $\pi$ bands saddle point, where the band-induced effective mass cannot be properly defined, and the contribution of the residual many-body interactions between electrons and holes should become important in "dressing" the quasiparticle. For these reasons, there are no theoretical predictions in the literature for such a case.

Nonetheless, given the proposed role of the hot-e$^{-}$ excitations to the IPS population processes, such a large BGR shift should imply a significant many-body contribution to the IPS properties, such as a deviation of the IPS effective mass and linewidth from the free-electron values.

\begin{figure}
\begin{center}
\includegraphics[width=0.9\linewidth, bb=55 110 518 570]{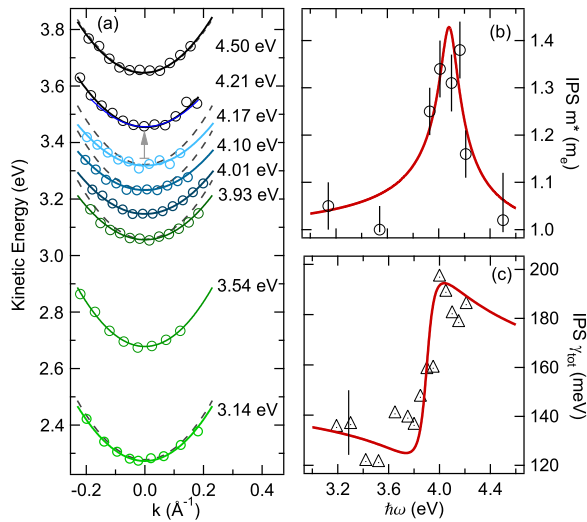}
\caption{(a) IPS $\bk_{||}$ dispersions at various photon energies (open circles). The parabolic fits (solid lines) are also shown. The free-electron parabola (dashed lines) is reported for comparison. The $\hbar\omega=4.21$ eV dispersion has been rigidly shifted (gray arrow) of about 0.1 eV upwards for clarity. (b) The IPS $m^*$ data extracted from dispersions are shown versus photon energy (open circles). The red solid line results from the fit of Eq. \ref{eq:mass}. (c) IPS PE linewidth versus photon energy (open triangles) from data reported in Fig. \ref{FIG2}(a). The red solid line is the fit of Eq. \ref{snug}.}
\label{FIG3}
\end{center}
\end{figure}

To verify this assumption, the IPS $m^*$ at photon energies between $\hbar\omega$=3.14 eV and $\hbar\omega$=4.50 eV have been measured from $\bk_{||}$ dispersions, according to standard procedures. The IPS dispersions are shown in Fig. \ref{FIG3}(a), while the resulting $m^*$ versus photon energy is reported in Fig. \ref{FIG3}(b). In Fig. \ref{FIG3}(c) the IPS linewidth at $\bk_{||}=0$ versus photon energy, obtained from the NE spectra of  Fig. \ref{FIG2}(a), is shown. The $m^*$ versus the photon energy shows a maximum located at $\hbar\omega\approx$4.1 eV, and corresponding to an increase of about 34$\%$ of the bare electron mass (m$_\mathrm{e}$), as measured at $\hbar\omega\approx$3.14 eV. 

Noteworthy, the effective mass rising is accompanied by a sudden increase of the IPS NE linewidth, occurring near 4 eV, from the reported value of $\gamma_0\approx130$ meV \cite{LehmannPRB1999} to $\approx$180 meV. These experimental evidences strongly point towards a relevant coupling between the surface IPS and the bulk optical excitations in graphite.

The IPS $m^*$ and linewidth dependence on photon energy can be rationalized on the basis of a semi-empirical self-energy formalism, by considering the interaction between the IPS and the photoinduced hot-e$^-$ gas in the $\pi^*$ band as the cause of the observed deviations from the free-electron case. Accordingly, a simple choice for the functional form of the real part of the IPS self-energy $\Sigma_\omega(\bk_{||})$ explaining the observed frequency-dependent dispersion modifications is the following,
\begin{equation}
\Re \Sigma_\omega(\bk_{||}) = N^2(\omega)\left(\alpha - \beta\frac{\hbar^2\bk_{||}^2}{2m_\mathrm{e}} \right)
\label{realsigma}
\end{equation}
where
\begin{equation}
N(\omega)= (1-R)F(\hbar c)^{-1}\epsilon_2(\omega-d\omega)
\label{nom}
\end{equation}
is the non-equilibrium photoexcited carrier density in the $\pi$ bands \cite{grosso}; here $d\omega$ is the renormalization energy related to the BGR shift. $\alpha$ and $\beta$ are two parameters of the model: $N^2(\omega)\alpha$ represents the rigid shift of the dispersion, whereas $N^2(\omega)\beta $ is the positive parameter representing the effective mass increase. 
From Eq. \ref{realsigma} is possible to obtain the $m^*$ dependence on the photon energy 
\begin{equation}
\label{eq:mass}
m^*(\omega) = m_\mathrm{e}\left[1-\beta N^2(\omega)\right]^{-1}.
\end{equation} 
The contribution to the IPS line-width at $\bk_{||}$=0 from the scattering with hot-e$^-$ can be obtained from the $\Im\Sigma_\omega(\bk_{||}=0)$ using Eq. \ref{realsigma} and the Kramers-Kronig (K-K) relations \cite{Lucarini}. The total linewidth is given by,
\begin{equation}
\gamma_\textrm{tot}(\omega) =  \gamma_0 + \alpha \Im \frac{2}{\pi}\int_{0}^{\infty}d\nu \frac{(\Gamma_0-i\omega)N^2(\nu)}{\nu^2 - (\omega + i\Gamma_0)^2},
\label{snug}
\end{equation}
where $\gamma_0$=130 meV is the equilibrium IPS linewidth, and $\Gamma_0=0.07$ eV is a Lorentzian broadening. Since experimental data on dielectric constant are affected by a substantial experimental broadening, calculated values of $\epsilon_2(\omega)$ \cite{PedersenPRB2003} has been used.  

Fig. \ref{FIG3}(b) shows the comparison of the fitting of Eq. (\ref{eq:mass}) (red curve) to the IPS $ \textrm{m}^*$ (open circles), using the coupling constant $\beta$ and the BRG shift $d\omega$ as parameters. Instead, the fit of Eq. \ref{snug} to the linewidth data is reported in Fig. \ref{FIG3}(c), using $\alpha$ and $d\omega$ as parameters. The best $d\omega$ fit value results 0.4 $\pm$ 0.1 eV from both $m^*$ and linewidth data. The fact that similar $d\omega$ values are obtained from two independent measurements, i.e. $m^*$ and $\gamma_{\textrm{tot}}$, connected by the K-K relations and that this $d\omega$ value is consistent with the energy shift between the IPS intensity and the $\epsilon_2(\omega)$ maximum as measured at equilibrium, i.e. $\approx0.4$ eV are considered as strong evidences of the photoinduced residual interactions between photoinduced excitations and image potential states in graphite.

Their strong interaction can be rationalized considering the low interlayer coupling between graphene layers in graphite that prevents the diffusion of photoexcited carriers in the bulk. Moreover, photoinduced excitations at the $\pi$-bands saddle point have already proven to strongly deviate from the Fermi-liquid behavior \cite{MoosPRL2001,SpataruPRL2001}], confirming the peculiar character of the M point of the BZ in graphite. Studying the residual interactions dependence on the photodoping concentration would be important in order to clarify the detailed properties of the many-body interactions. This would require performing fluence-dependent measurements over a substantial (i.e. at least a decade) fluence range. Unfortunately two concurring causes such as space charge effects and low electron statistics, limit the dynamic range of these experiments. However, it is possible to finely tune the absorbed fluence (and hence the $\pi$ carrier concentration) by varying the manipulator angle employing a non-symmetric experimental geometry, in which the plane of incidence of light is orthogonal to the manipulator axis; this induces angle-dependent variations of the absorbed fluence, resulting in a $\bk_{||}$-dependence in the measured IPS effective mass, in accordance with Eq. \ref{eq:mass}. \cite{Pagliara_unp}.   

In conclusion, we have reported the experimental evidence of the interactions between the photoinduced excitations and the image potential states (IPS) on highly oriented pyrolitic graphite (HOPG) when photoexcited by an intense (at F $\approx100$ $\mu$J cm$^{-2}$), ultrashort ( $\approx100$ fs) coherent light pulses, in the 3.1- 4.5 eV photon energy range. The strong dependence on the photon energy of the IPS photoemission intensity, effective mass, and linewidth, observed by NL-ARPES, when $\hbar\omega$ approaches the $\pi\rightarrow\pi^*$ saddle points absorption has been attributed to many-body effects (residual interactions) originating from the high density of excitations. Finally, these findings bring a clear evidence of a high IPS-bulk coupling in graphite and opens the way for exploiting the IPS as a sensitive, nonperturbing probe for the many-body dynamics in materials.

\bibliography{articolo,books}
\end{document}